\documentstyle[prl,aps,mypsfig2,twocolumn]{revtex}

\begin{document}

\noindent {\Large \sf Implementation of a three-quantum-bit\\ search algorithm}\\

\noindent{\large \sf Lieven M.K. Vandersypen and Matthias Steffen}\\
Solid State  and Photonics Laboratory, Stanford University,  Stanford, 
CA 94305-4075\\
IBM Almaden Research Center, San Jose, CA 95120\\
\noindent{\large \sf Mark H. Sherwood, Costantino S. Yannoni, Gregory Breyta and Isaac L. Chuang}\\
IBM Almaden Research Center, San Jose, CA 95120\\

\def\be{\begin{equation}}
\def\ee{\end{equation}}
\newcommand{\ket}[1]{\mbox{$|#1\rangle$}}
\newcommand{\mypsfig}[2]{\psfig{file=#1,#2}}

\vspace*{-0.3cm}

\begin{abstract}

We report the experimental implementation of Grover's quantum search algorithm on a quantum computer with three quantum bits.  The computer consists of molecules of $^{13}$C-labeled CHFBr$_2$, in which the three weakly coupled spin-1/2 nuclei behave as the bits and are initialized, manipulated, and read out using magnetic resonance techniques. This quantum computation is made possible by the introduction of two techniques which significantly reduce the complexity of the experiment and by the surprising degree of cancellation of systematic errors which have previously limited the total possible number of quantum gates.
\end{abstract}

\vspace*{0.5cm}

Elementary quantum computations~\cite{qc-idea,shor,grover} have recently
been demonstrated experimentally~\cite{nmrqc-algorithms-demos} using coupled
nuclear spins as quantum bits (qubits) and solution nuclear magnetic
resonance (NMR) techniques to prepare, manipulate, and detect the
spins~\cite{nmrqc-idea,ernst-freeman}.  Aside from the well-understood
scaling limitations due to the use of a high-temperature (almost random)
system instead of a low-temperature (low entropy) polarized spin
system~\cite{nmrqc-idea,labeling}, the crucial limitation in applying this method
to implement larger quantum algorithms has been systematic errors in the
quantum gates.  These gates are implemented by applying pulses of
radiofrequency (RF) electromagnetic fields of precise duration and phase,
which are in practice highly inhomogeneous over the sample volume,
causing the gate fidelity\cite{nielsen} to be less than $95$\%.
Producing a homogeneous field is difficult because of the 
sample geometry and the necessity of keeping the field transverse to the
$11.8$ Tesla alignment field.  If such systematic errors 
simply accumulated, these observations would imply that fewer than 90 gates~\cite{max_gates} applied to any one spin could ever be cascaded in these systems.

One technique which has been proposed for controlling errors in quantum
gates is quantum error correction\cite{q-error-corr}, and in fact, it is known that
if the error probability per gate is smaller than a certain
threshold~\cite{accur-threshold}, and in addition the architecture of the
computer satisfies certain criteria, then quantum computation could be
carried out robustly and indefinitely, despite the errors.  Unfortunately,
this threshold is currently estimated to be below $10^{-4}$, which is virtually
unreachable with current NMR quantum computation techniques and present
levels of systematic errors.
Now, the principle attribute of traditional error correction
techniques is their ability to correct completely random errors which
originate from fundamentally irreversible {\em decoherence}
phenomena~\cite{decoherence}; in principle, systematic errors, which are inherently
{\em reversible} --- at least on an appropriate time scale --- should be much
easier to control, given knowledge about their origin.

Here, we report on the experimental realization of a three-qubit quantum
search algorithm, and the observation of {\em 28} full cycles of the
theoretically predicted oscillatory behavior for the algorithm.
This was made possible by the application of two techniques to reduce the number of gates required for the implementation, and by the control and cancellation of systematic errors. More than 280 two-qubit quantum gates involving 1350 RF pulses were successfully cascaded, which exceeds not only the number of gates used in all previous NMR quantum computing experiments but also the limitation of 90 gates, imposed by cumulative systematic errors.

%%%%%%%%%%%%%%%%%%%%%%%%%%%%%%%%%%%%%%%%%%%%%%%%%%%%%%%%%%%%%%%%%%%%%%%%%%%%%

The experiments were carried out with molecules containing $n=3$ weakly
coupled spin-1/2 nuclei, subject to a strong magnetic field
$\overrightarrow{B}_0$. Each spin represents one qubit, with the ground and
excited state serving as $\ket{0}$ and $\ket{1}$. The Hamiltonian is well
approximated by ($\hbar=1$)~\cite{ernst-freeman}
\be 
{\cal H} = - \sum_{i} \omega_i I_{z i}
 	   + \sum_{i<j} 2\pi J_{i j} I_{z i} I_{z j}
	   + {\cal H}_{env}
\label{eq:hamiltonian}
\ee
where the first term represents the free precession of each spin $i$ about
$-\overrightarrow{B}_0$, with $I_{z i}$ the angular momentum operator in the
$\hat{z}$ direction and $\omega_i$ the Larmor frequency of spin $i$. The
second term describes a scalar spin-spin coupling of $J_{i j}$ between spins
$i$ and $j$ and ${\cal H}_{env}$ represents coupling to the environment,
which causes decoherence. A resonant RF field rotating in the
$\hat{x}$-$\hat{y}$ plane can be gated on to perform single-spin rotations.

%%%%%%%%%%%%%%%%%%%%%%%%%%%%%%%%%%%%%%%%%%%%%%%%%%%%%%%%%%%%%%%%%%%%%%%%%%%%%

\begin{figure}[htbp]
\begin{center}
\mbox{\psfig{file=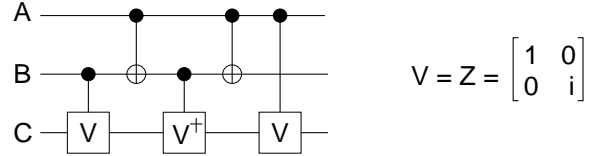,width=3in}}\\
\end{center}
\caption{ \narrowtext Quantum circuit for a three-qubit Toffoli gate which
flips the phase of the $\ket{1} \ket{1} \ket{1}$ term. Vertical lines
connecting two horizontal lines represent two-qubit gates, which perform the
unitary operation shown on the target qubit if and only if the control qubit
(indicated by a black dot) is $\ket{1}$ (an open circle represents a 
NOT-gate). Time goes from left to right.}
\label{fig:toffoli}
\end{figure}

The first method to reduce the number of one- and two-spin gates
used to realize any given $n$-qubit unitary operation
starts with a library of efficient implementations for often-used building
blocks. For example, four equivalent realizations of a $Z$-rotation using an RF coil in the transverse plane, are $Z \equiv Y X \bar{Y} \equiv \bar{Y} \bar{X} Y \equiv \bar{X} Y X \equiv X \bar{Y} \bar{X}$, where $X$ and $\bar{X}$ denote +90$^{\circ}$ and -90$^{\circ}$ rotations about the $\hat{x}$ axis. 
And CNOT$_{i j}$ 
$\equiv X_j Y_j \tau_{i j} Z_i \bar{Y}_j
\equiv \bar{X}_j \bar{Y}_j \tau_{ij} \bar{Z}_i Y_j
\equiv Y_j \tau_{ij} Z_i \bar{Y}_j X_j
\equiv \bar{Y}_j \tau_{ij} \bar{Z}_i Y_j \bar{X}_j$
where $\tau_{i j}$ describes a time evolution of $1/2J_{i j}$, during which all couplings except $J_{i j}$ are refocused~\cite{nmrqc-idea} (CNOT$_{i j}$ represents a controlled-NOT which flips spin $j$ if and only if the control spin $i$ is $\ket{1}$).
For any given quantum circuit, e.g. Fig.~\ref{fig:toffoli} (after~\cite{barenco}), specific implementations of the building blocks can then be laid out such that adjacent pulses invert each other, so they can be omitted.
The simplification process can exploit the commutation of $Z$ rotations with time evolutions (the Hamiltonian), and of RF pulses on different spins with each other. Furthermore, NMR permits a direct implementation of many useful gates such as the controlled-$V$ (Fig.~\ref{fig:toffoli}), which can be realized by $\tau_{i j}/2$ and a few single-spin rotations. If only CNOT's and 1-qubit gates were used to implement the controlled-$V$'s (as in~\cite{barenco}) and if no pulses were canceled out, 70.5 $\pi/2$ pulses and 8 evolutions of $1/2J$ would be required for the Toffoli gate (Fig.~\ref{fig:toffoli}). This was reduced to 19 pulses, 2 evolutions of $1/2J$ and 3 of $1/4J$, by using the simplification methodology~\cite{pulse-sequence}.

%%%%%%%%%%%%%%%%%%%%%%%%%%%%%%%%%%%%%%%%%%%%%%%%%%%%%%%%%%%%%%%%%%%%%%%%%%%%

The second method to reduce the complexity concerns the initialization of the qubits to the ground state, generally the first step in quantum algorithms. The density matrix $\rho_{eq}$ of the system of Eq.~\ref{eq:hamiltonian} in equilibrium at room temperature is highly mixed ($\hbar \omega_i \ll k_B T$, with $k_B T$ the thermal energy).
The use of room temperature nuclear spins is made possible by converting the equilibrium state into an ``effective pure state''~\cite{nmrqc-idea} (one population deviates from a uniform background). We propose the preparation of an effective pure ground state $\rho_{g}$ by a new variant of temporal averaging~\cite{temp-lab}. The original scheme involves the summation of $2^n-1$ density matrices, each of which is obtained from $\rho_{eq}$ by cyclicly permuting all populations except the ground state population~\cite{temp-lab}; but this gets rather involved for $n \geq 3$. A much simpler approach is to take a weighted sum of (diagonal) density matrices $\rho_l$ which are obtained from $\rho_{eq}$ by performing just a few CNOT's to rearrange the $2^n-1$ populations.
The weights $w_l$ are determined by solving a set of linear equations $\sum_l w_l \, \mbox{diag}(\rho_l) = \mbox{diag}(\rho_{g})$. This approach may still require up to $2^n-1$ experiments to get exactly $\rho_{g}$, but allows one also to approximate $\rho_{g}$ using far fewer experiments.
All the data shown below were obtained using just three experiments, resulting in an expected variance of the $2^n-1$ populations of only 7$\%$ of their average value.
These modifications thus increase both the simplicity and scalability of temporal labeling.

%%%%%%%%%%%%%%%%%%%%%%%%%%%%%%%%%%%%%%%%%%%%%%%%%%%%%%%%%%%%%%%%%%%%%%%%%%%%%

We used these methods to implement the eight instances of the 3-qubit Grover algorithm. This algorithm achieves a quadratic speed-up over classical algorithms when searching in an unsorted database with $N$ entries~\cite{grover}: given an unknown function $f(x)$ with the promise that $f(x_0)=-1$ for a unique $x_0$, and $f(x)=1$ elsewhere, a classical search needs ${\cal O}(N)$ (4.375 for $N$=8) attempts to find $x_0$, while ${\cal O}(\sqrt{N})$ (2 for $N$=8) queries suffice using Grover's quantum algorithm~\cite{grover}.
For $N$=8, the algorithm requires four Toffoli gates~\cite{barenco}, plus several 1-qubit gates. Earlier experiments~\cite{nmrqc-algorithms-demos,labeling} required at most one Toffoli gate and a few two-qubit gates.

%%%%%%%%%%%%%%%%%%%%%%%%%%%%%%%%%%%%%%%%%%%%%%%%%%%%%%%%%%%%%%%%%%%%%%%%%%%%%

\begin{figure}[h]
\mbox{\psfig{file=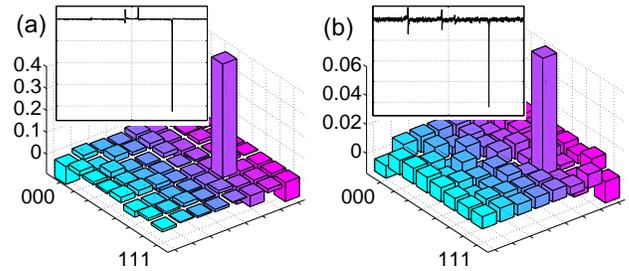,width=3.4in}}\\
\caption{Experimental deviation density matrices $\rho_{exp}$ for $\ket{x_0}=\ket{1} \ket{0} \ket{1}$, shown in magnitude with the sign of the real part (all imaginary components were small), after 2 {\bf (a)} and 28 {\bf (b)} Grover iterations. The diagonal elements give the population difference with respect to the average. The off-diagonal elements represent coherences between the basis states. {\sf Inset}: The corresponding $^{13}$C spectra ($^{13}$C was the least significant qubit). The receiver phase and read-out pulse are set such that the spectrum be absorptive and positive for a spin in $\ket{0}$.}
\label{fig:denmats}
\end{figure}

We selected $^{13}$C-labeled CHFBr$_2$~\cite{synthesis} for our
experiments. The Hamiltonian of the $^1$H-$^{19}$F-$^{13}$C spin system is
of the form of Eq.~\ref{eq:hamiltonian} with $J_{HC}$ = 224 Hz, $J_{HF}$ =
50 Hz and $J_{FC}$ = $-311$ Hz. The scalar interaction with the Br nuclei is
averaged out and only contributes to ${\cal
H}_{env}$~\cite{ernst-freeman}. Experiments were carried out at IBM using an
11.7 T Oxford Instruments magnet and a Varian Unity Inova spectrometer with
a triple resonance (H-F-X) probe from Nalorac. The output state of each spin
$i$, $\ket{0}$ or $\ket{1}$, can be determined from the phase of the signal
induced in the RF coil after applying an $X_i$ read-out pulse. In fact, the
spectrum of the signal of any one spin suffices to determine the output
state of all the spins given that they are in an effective pure energy
eigenstate and that they are all mutually coupled: each spectrum then
contains only a single line, the frequency of which, combined with the
knowledge of the $J_{i j}$, reveals the state of the remaining spins
(Fig.~\ref{fig:denmats} (a), inset).  We also reconstructed the complete
output {\em deviation} (traceless) density matrix~\cite{nmrqc-idea} using
quantum state tomography~\cite{tomography} (Fig.~\ref{fig:denmats} (a)).

%%%%%%%%%%%%%%%%%%%%%%%%%%%%%%%%%%%%%%%%%%%%%%%%%%%%%%%%%%%%%%%%%%%%%%%%%%%%%%

\begin{figure}[h]
\mbox{\psfig{file=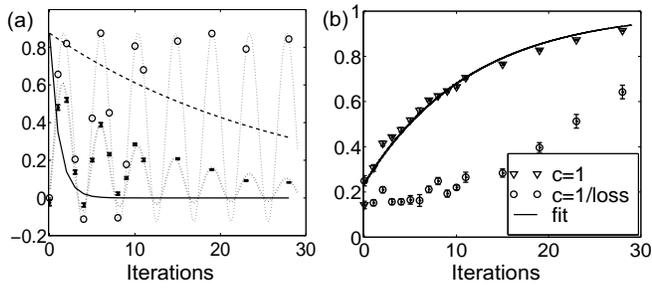,width=3.4in}}\\
\caption{{\bf (a).} Experimental (error bars) and ideal (circles) amplitude of $d_{x_0}$, with fits (dotted) to guide the eye. Dashed line: the signal decay for $^{13}$C due to intrinsic phase randomization or decoherence (for $^{13}$C, $T_2 \approx 0.65$ s). Solid line: the signal strength retained after applying a continuous RF pulse of the same cumulative duration per Grover iteration as the pulses in the Grover sequence (averaged over the three spins; measured up to 4 iterations and then extrapolated). {\bf (b)}. The relative error $\epsilon_r$.}
\label{fig:error-iter}
\end{figure}

The agreement between experimental results and theoretical predictions is good, considering that about 100 pulses were used and that the systematic error rate exceeds $5\%$ per RF pulse (the measured signal loss due to RF field inhomogeneity after applying $X_i$). This suggests that the systematic errors cancel each other out to some degree. We examined this in more detail in a series of experiments with increasingly longer pulse sequences executing up to 28 Grover iterations (repeated executions of the two main steps of Grover's algorithm~\cite{grover}). Theoretically, the probability of $\ket{x_0}$ oscillates as a function of the number of iterations $k$, reaching a first maximum for $k = {\cal O}(\sqrt{N})$~\cite{grover}. Fig.~\ref{fig:error-iter} (a) shows that the diagonal entry $d_{x_0}$ of $\rho_{exp}$ oscillates as predicted but the oscillation is damped as a result of errors, with a time constant $T_d$ of 12.8 iterations. 
However, $T_d$ would have been smaller than 1.5 if the errors due to just the RF field inhomogeneity were cumulative (Fig.~\ref{fig:error-iter} (a), solid line). 
Remarkably, after a considerable initial loss, $d_{x_0}$ decays at a rate close to the $^{13}$C $T_2$ decay rate~\cite{ernst-freeman} (dashed line), which can be regarded as a lower bound on the overall error rate.
A more complete measure to quantify the error and benchmark results is the relative error $\epsilon_r = \parallel c \rho_{exp} - \rho_{th} \parallel_2 / \parallel \rho_{th} \parallel_2$, where $\rho_{exp}$ and $\rho_{th}$ are the experimental and theoretical (traceless) deviation density matrices, and $\parallel \cdot \parallel_2$ is the 2-norm ~\cite{2-norm}. Comparison of $\epsilon_r$ with $c=1$ and $c$ equal to the inverse of the signal loss (Fig.~\ref{fig:error-iter} (b)) reveals that signal loss dominates over other types of error. Furthermore, the small values of $\epsilon_r$ with $c>1$ suggest that $\ket{x_0}$ can be unambiguously identified, even after almost 1350 pulses. This is confirmed by the density matrix measured after 28 iterations, which has a surprisingly good signature (Fig.~\ref{fig:denmats} (b)). Given the error of $>5\%$ per single $\pi/2$ rotation, all these observations demonstrate that substantial cancellation of errors took place in our experiments.

%%%%%%%%%%%%%%%%%%%%%%%%%%%%%%%%%%%%%%%%%%%%%%%%%%%%%%%%%%%%%%%%%%%%%%%

The error cancellation achieved was partly due to a judicious choice of the
phases of the refocusing pulses, but a detailed mathematical description in
terms of error operators is needed to fully exploit this effect in arbitrary
pulse sequences. This difficult undertaking is made worthwhile by our
observations. This conclusion is strengthened by a similar observation
in~\cite{labeling}. Also, we believe that error cancellation behavior is
{\em not} a property of the Grover iterations, because we found
experimentally that the choice of implementation of the building blocks
dramatically affects the cancellation effectiveness. Whereas the
cancellation of systematic errors makes it possible to perform surprisingly
many operations,
the methods for simplifying pulse sequences reduce the number of operations
needed to implement a given quantum circuit. This combination permitted the 
succesful realization of Grover's algorithm with 3 spins and brings many
other interesting quantum computing experiments within reach.

%%%%%%%%%%%%%%%%%%%%%%%%%%%%%%%%%%%%%%%%%%%%%%%%%%%%%%%%%%%%%%%%%%%%%%%

We thank J. Harris, W. Risk and A. Pines for their support. L.V. 
acknowledges a Yansouni Family Stanford Graduate Fellowship.  This work was
performed under the auspices of the DARPA NMRQC initiative.

%%%%%%%%%%%%%%%%%%%%%%%%%%%%%%%%%%%%%%%%%%%%%%%%%%%%%%%%%%%%%%%%%%%%%%%

\begin{thebibliography}{30}
\bibitem{pulse-sequence} The detailed pulse sequence is available from the authors.
\bibitem{synthesis} Synthesized by heating a mixture of $^{13}$CHBr$_3$ (2.25 g, CIL) and HgF$_2$ (2.8 g, Aldrich) in increments (5$^{\circ}$C for 15 min) from 70$^{\circ}$C to 85$^{\circ}$C in a Kugelrohr apparatus and condensing the product into a cooled bulb.  This material was re-distilled bulb-to-bulb at 65$^{\circ}$C to give 750 mg (99 $\%$ purity) of $^{13}$CHFBr$_2$, which was dissolved in d6-acetone.
\bibitem{2-norm} The 2-norm gives the absolute value of the largest eigenvalue. It is a pessimistic measure, compared to the traditional $1- \mbox{Tr}(\sqrt{\rho_1} \rho_2 \sqrt{\rho_1})$ (the latter is defined only for non-negative matrices, though).
\end{thebibliography}
\end{document}